# Quality-Driven Disorder Handling for M-way Sliding Window Stream Joins

Yuanzhen Ji*‡, Jun Sun*‡, Anisoara Nica†, Zbigniew Jerzak*, Gregor Hackenbroich*, Christof Fetzer‡
*SAP SE, Germany  †SAP SE, Canada
Email: firstname.lastname@sap.com
‡Systems Engineering Group, Technische Universität Dresden, Germany
Email: christof.fetzer@tu-dresden.de

*Abstract*—Sliding window join is one of the most important operators for stream applications. To produce high quality join results, a stream processing system must deal with the ubiquitous disorder within input streams which is caused by network delay, asynchronous source clocks, etc. Disorder handling involves an inevitable tradeoff between the latency and the quality of produced join results. To meet different requirements of stream applications, it is desirable to provide a user-configurable result-latency vs. result-quality tradeoff. Existing disorder handling approaches either do not provide such configurability, or support only user-specified latency constraints.

In this work, we advocate the idea of *quality-driven* disorder handling, and propose a buffer-based disorder handling approach for sliding window joins, which minimizes sizes of input-sorting buffers, thus the result latency, while respecting user-specified result-quality requirements. The core of our approach is an analytical model which directly captures the relationship between sizes of input buffers and the produced result quality. Our approach is generic. It supports $m$-way sliding window joins with arbitrary join conditions. Experiments on real-world and synthetic datasets show that, compared to the state of the art, our approach can reduce the result latency incurred by disorder handling by up to $95\%$ while providing the same level of result quality.

## I. INTRODUCTION

In the past two decades, we have witnessed an increasing interest in real-time processing of data streams generated by online sources such as sensor networks [1] and financial markets [2]. A stream tuple is often tagged with a timestamp as it is generated at a data source, e.g., a sensor. For many stream applications, $m$-way sliding window joins (MSWJ) are key operators for discovering correlations across different streams, e.g., finding similar news items from different news sources [3]. In general, an MSWJ operator works as follows [4]–[6]: for each newly arrived tuple $e$ from any input stream, the operator invalidates expired tuples in windows on all other input streams based on the timestamp of $e$, probes all remaining tuples in those windows, and produces derived result tuples that satisfy the predefined join condition $p^{\bowtie}$. The timestamp assigned to a result tuple is the maximum timestamp among its deriving input tuples.

One big challenge in data stream processing is to handle *stream disorder* with respect to tuple timestamps. For multi-input operators such as MSWJ, stream disorder enters in two forms: First, each individual stream could be out of order; namely, tuples within the stream do not arrive in non-decreasing timestamp order (*intra-stream disorder*). Second, even if each individual stream is timestamp-ordered, tuples from different streams could arrive out of order (*inter-stream disorder*). Disorder is ubiquitous in real-world data streams transmitted across a network because of the inherent network asynchrony. For instance, data sent to a "sink" node in a sensor network may experience different delays and hence arrive at the sink node in an arbitrary order.

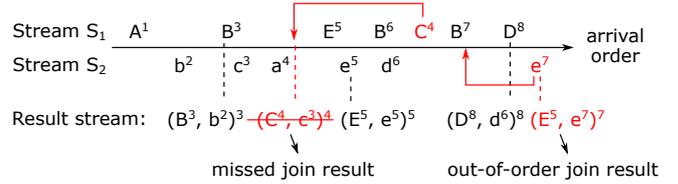

Fig. 1. Effects of stream disorder on join results (window size $W_1=W_2=2$)

Stream disorder must be handled carefully; otherwise, it will cause degradation of the result quality. Specifically, we would miss valid join results or produce out-of-order results. Fig. 1 illustrates these effects of stream disorder on join results. For simplicity, we consider a 2-way join $S_1 \bowtie S_2$. A sliding window of 2 time unit is applied on each stream. An input tuple is represented by $x^{ts}$, where $x$ is the value of the tuple's join attribute, and $ts$ is its timestamp. To differentiate between tuples from $S_1$ and $S_2$, we use capital letters to represent attribute values for $S_1$ tuples. The timestamp of a result tuple is also denoted by superscript. In this example, tuple $C^4$ is an out-of-order tuple in $S_1$. Without disorder handling, we would miss the result tuple $(C^4, c^3)^4$ that can be derived from $C^4$; because by the time $C^4$ arrives, its matching tuple $c^3$ in $S_2$ has expired from the current window on $S_2$ because of the previous arrival of $B^6$. Stream $S_2$ is timestamp-ordered; however, the tuple $e^7$ in $S_2$ arrives after the tuple $D^8$ in $S_1$, and hence is out of order from the perspective of the join operator. Although in this case, the result tuple $(E^5, e^7)^7$ can still be produced[1], it is out of order in the result stream. Out-of-order output might be unacceptable in many scenarios, e.g., when the output is consumed for feedback control. In this work, we focus on these applications and hence target for an in-order output stream.

To avoid missed and out-of-order join results, we need to either sort input tuples to process them in a timestamp order [9], [10], or sort result tuples [10], [11]. In either case, the latency of the results is increased. This implies that high result quality and low result latency are two conflicting targets in disorder handling. Many applications, such as network traffic monitoring and environment monitoring, allow doing incomplete sorting and hence losing a small fraction of join results to obtain low result latency[2]. However, it is still desired that the result quality is controlled at an acceptable level.

---

[1]Note that some join algorithms (e.g., [7], [8]) would miss the result tuple $(E^5, e^7)^7$; because upon receiving a new tuple, they invalidate expired tuples in windows on all streams. Hence, at the arrival of $D^8$, the tuple $E^5$ would expire from the window on $S_1$.

Based on observations that (1) disorder handling involves an inevitable tradeoff between the result latency and the result quality, and (2) different applications may have different tradeoff requirements, we believe that a disorder handling approach should support user/application-configurable result-latency vs. result-quality tradeoff. Existing disorder handling approaches either do not provide this configurability (e.g., [12], [13]), or support disorder handling only under user-specified latency constraints (e.g., [14], [15]). As a complement to the state of the art, in our previous work [16], [17], we proposed the idea of *quality-driven disorder handling*, with the objective of *minimizing the result latency while honoring user-specified result-quality requirements*. In [16], [17], we introduced an adaptive, quality-driven, buffer-based disorder handling approach for processing sliding window aggregate queries. The size of the buffer for sorting input tuples was dynamically adjusted using a proportional-derivative (PD) controller.

In this paper, we extend our focus beyond unary stream processing operators to n-ary operators, in particular, the MSWJ operator. In contrast to the existing work such as [15] and [18], we assume the most common scenario, where no facilitating information such as punctuations is available in input streams to indicate the stream progress; hence, we again employ the buffer-based disorder handling approach. However, due to the difference in the operator semantics, our result-quality metric for joins is different from that for aggregates. From the discussion above, we can see that the output of an MSWJ under incomplete disorder handling is only a subset of the output produced when the input streams are in order and synchronized with each other. Hence, we adopt the metric *recall*, which is often used in work about load shedding for joins [3], [8], [19]. In addition, the PD controller, which was used in our previous work for adapting the size of the input buffer, treats the relationship between the applied buffer size and the consequent result quality as a black box; as a result, in one adaptation step, it is not able to directly determine the optimal buffer size for meeting the result-quality requirement. In this paper, we propose a new adaptation method, which allows searching for the optimal buffer size directly at one adaptation step, by analytically modeling the relationship between the applied buffer size and the produced result quality. Moreover, the contribution of an input tuple to the quality of join results is determined by its *productivity*, i.e., the number of join results that it can derive. Clearly, the unsuccessful handling of a delayed tuple with a high productivity has a higher impact on the result quality than the unsuccessful handling of a delayed tuple with a low productivity. Hence, the proposed analytical model also considers the correlation between the delay and the productivity of tuples.

In summary, our work makes the following contributions:
- We extend the application of the idea of quality-driven disorder handling to MSWJ queries executed over out-of-order and unsynchronized data streams. The objective is to minimize the result latency incurred by disorder handling while respecting the user-specified result-quality requirement. To this end, we propose a generic buffer-based disorder handling approach which supports MSWJs with arbitrary join conditions. (Sec. III)
- We show that it suffices to use a buffer of the same size on each input stream to deal with its intra-stream disorder, which may sound non-intuitive since different streams normally have different disorder characteristics. (Sec. III-B)
- We analytically model the relationship between the applied buffer sizes and the consequent quality (i.e., recall) of produced join results. Based on this model, we propose a new buffer-size adaptation method, which can search for the optimal buffer size to meet the user-specified result-quality requirement at each adaptation step. (Sec. IV-A)
- We incorporate the consideration of potential correlation between the delay and the productivity of tuples into the proposed analytical model. Inspired by the work of [3], [20], we learn this correlation at query runtime by profiling the output of the join. This method is light-weight and can work with arbitrary join conditions. (Sec. IV-B)

We introduce related concepts and notations in Sec. II. We discuss how the proposed approach can be applied in the context of distributed MSWJ processing in Sec. V. We conduct experiments on both real-world and synthetic datasets to study the effectiveness of our solution. The results are reported in Sec. VI. We review related work in Sec. VII and conclude the paper in Sec. VIII.

## II. PRELIMINARIES

### A. MSWJ over Data Streams

An MSWJ has $m \geq 2$ input streams $S_1, S_2, \ldots, S_m$, and an optional join condition $p^{\bowtie}$ which may contain one or more join predicates. We denote the $j$-th arrived tuple in stream $S_i$ by $e_{i,j}$, and its timestamp by $e_{i,j}.ts$. We define the *local current time* $^iT$ of stream $S_i$ as the maximum timestamp among already arrived tuples in $S_i$, i.e., $^iT = max\{e_{i,j}.ts | e_{i,j} \in S_i\}$[3]. A stream $S_i$ is considered to be *out of order* if it contains tuples $e_{i,j}$ and $e_{i,k}$ such that $j < k$ and $e_{i,j}.ts > e_{i,k}.ts$. The tuple $e_{i,k}$ in this case is called an *out-of-order tuple* in $S_i$. For a tuple $e_i$ in any stream $S_i$, the *delay* of the tuple is denoted by $delay(e_i)$, and is defined as the difference between the $^iT$ updated at the arrival of $e_i$ and the timestamp of $e_i$, i.e., $delay(e_i) = {^iT} - e_i.ts$. We assume that the delay of any input tuple is bounded, but do not require knowing this upper bound a priori. At the query runtime, we take the maximum observed tuple delay as an estimation of this upper bound. In the following, we omit the subscript of a tuple completely or keep only the part that indicates the input stream, if the omitted part is not important in the context of the discussion.

We denote the *time skew* between a pair of streams $S_i$ and $S_j$ ($i \neq j$) by $skew(S_i, S_j)$, and define it as the absolute difference between the local current time of $S_i$ and $S_j$, i.e., $skew(S_i, S_j) = |{^iT} - {^jT}|$. As $^iT$ and $^jT$ are updated by newly arrived tuples in $S_i$ and $S_j$ respectively, $skew(S_i, S_j)$ often varies during the lifetime of $S_i$ and $S_j$. Furthermore, given $m$ streams, we refer to the stream with the smallest $T$ as the *slowest* stream in terms of the timestamp progress.

Each input stream $S_i$ is associated with a sliding window [21], whose size is specified by the user. In this paper, we focus on time-based sliding windows, where the window size $W_i$ is defined in the number of time units. Each input stream could have a different window size. Semantically, an input tuple $e_i$ from any stream $S_i$ joins with the subset $\{e_j \mid e_i.ts - W_j \leq e_j.ts \leq e_i.ts + W_j\}$ of each stream $S_j$ ($j \neq i$). A result tuple $\langle e_1, e_2, \ldots, e_m \rangle$ is produced if the join condition $p^{\bowtie}$ is passed. The timestamp assigned to the result tuple is $max\{e_i.ts \mid i \in [1, m]\}$. Finding the optimal join order is orthogonal to our focus, and any existing work in this area (e.g., [5], [20]) can be applied.

---
[2]Incomplete sorting of result tuples leads to loss of join results as well, because result tuples that are still out of order after sorting are discarded to fulfill the "in-order output" requirement.

[3]For brevity, we use the superscript $i$ before $T$ to denote a property of input tuples.

TABLE I. LIST OF NOTATIONS

| Notations | Description |
|---|---|
| $S_i$ | $i$-th input stream |
| $e_{i,j}$ | $j$-th arrived tuple in $S_i$ |
| $e_{i,j}.ts$ | Timestamp of tuple $e_{i,j}$ |
| $W_i$ | User-specified sliding window size on stream $S_i$ |
| $^iT$ | Local current time of $S_i$; $^iT = max\{e_{i,j}.ts \mid e_{i,j} \in S_i\}$ |
| $skew(S_i, S_j)$ | Time skew between $S_i$ and $S_j$; $skew(S_i, S_j) = \mid ^iT - {}^jT \mid$ |
| $delay(e_{i,j})$ | Delay of tuple $e_{i,j}$; $delay(e_{i,j}) = {}^iT - e_{i,j}.ts$ |
| $P$ | User-specified result-quality measurement period |
| $\gamma(P)$ | Recall of result tuples of an MSWJ measured based on $P$ |
| $\Gamma$ | User-specified requirement on $\gamma(P)$ |
| $K_i$ | Dynamic size of the buffer in the $K$-slack component for stream $S_i$ |
| $T^{sync}$ | Current maximum timestamp among tuples released from *Synchronizer* |
| $^{\bowtie}T$ | Current maximum timestamp among tuples received by the join operator |
| $K_i^{sync}$ | Implicit buffer size in the *Synchronizer* that contributes to handling the intra-stream disorder of $S_i$ |
| $r_i$ | Data arrival rate of $S_i$ |
| $L$ | Adaptation interval applied in the *Buffer-Size Manager* (system param.) |
| $b$ | Size of a basic window (system param.) |
| $g$ | $K$-search granularity in one adaptation step (system param.) |
| $w_j^l, \mid w_j^l \mid$ | The $l$-th basic window on stream $S_j$, and cardinality of $w_j^l$, respectively |
| $\gamma(K, L)$ | Estimated recall of results produced within $L$ under a give value of $K$ |
| $\Gamma'$ | Requirement on $\gamma(K, L)$; derived from user-specified $\Gamma$ |
| $N_{prod}^{\bowtie}(K, L)$ | Join result size within $L$ under given configuration of $K$ |
| $N_{true}^{\bowtie}(L)$ | True join result size within $L$ if streams are in-order and synchronized |
| $N_{prod}^{\times}(K, L)$ | Cross-join result size corresponding to $N_{prod}^{\bowtie}(K, L)$ |
| $N_{true}^{\times}(L)$ | True cross-join result size corresponding to $N_{true}^{\bowtie}(L)$ |
| $D_i, D_i^K$ | Random variables representing the coarse-grained tuple delay in $S_i$ and in the corresponding stream of $S_i$ received by join operator, respectively |
| $f_{D_i}, f_{D_i^K}$ | Probability density functions of $D_i$ and $D_i^K$, respectively |

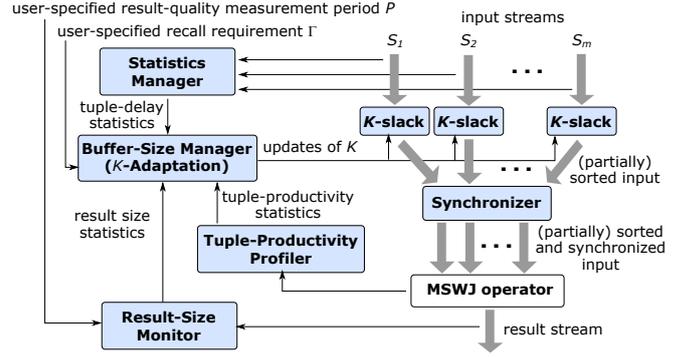

Fig. 2. Quality-driven disorder handling framework for MSWJs.

To guarantee an ordered output stream, we adopt the strategy of enforcing timestamp-ordered processing of tuples from all input streams [9], [10]; namely, the disorder handling is performed prior to the actual join operation. If the disorder handling is complete, then the join output (in terms of both the set of result tuples and the order between the result tuples with respect to their timestamps[4]) would be the same as the join output when input streams are totally in order and synchronized with each other. We define the number of result tuples produced *at the arrival* of a tuple $e$ in this case as the *productivity* of $e$. Details about our disorder handling solution are given in Sec. III and Sec. IV.

### B. Result-Quality Metric

When the disorder handling is incomplete, only a fraction of true results (i.e., results produced when streams are in order and synchronized) will be produced. The fraction of actually produced results is defined as the *recall* [8], and is used as the result-quality metric for MSWJs. In our work, each time the recall of results needs to be measured, instead of considering all join results produced so far, we consider the join results whose timestamps are within the last $P$ time units; $P$ is termed the *result-quality measurement period*, and it is a user-specified requirement. We introduce $P$ for two reasons: (1) it allows a user to specify the quality measurement period that is of his own interest; (2) with a full-history-based recall definition, it could happen that the fraction of produced results keeps high within a long period $I_1$ and keeps low within the following long period $I_2$, but the overall fraction of produced results within $I_1 + I_2$ still looks good. This may be undesirable for applications that favor continuous high result quality. With a period-based recall definition, the above described situation is detectable if the length of the period is set small. In this regard, a period-based recall is indeed a stricter quality metric compared to a full-history-based recall. Using the period-based recall definition, we can guide our disorder handling approach to provide a continuous high quality.

---
[4]We do not restrict the order among result tuples having the same timestamp.

Formally, given a user-specified result-quality measurement period $P$, we denote the recall of MSWJ results measured at any time with respect to $P$ by $\gamma(P)$, and define it as

$$\gamma(P) = \frac{\text{\# results whose } ts \text{ are within the last } P \text{ time units}}{\text{\# true results whose } ts \text{ are within the last } P \text{ time units}}.$$

To support quality-driven disorder handling, we allow a user to specify his requirement on the minimum achieved $\gamma(P)$. We denote this requirement by $\Gamma$.

## III. QUALITY-DRIVEN DISORDER HANDLING FRAMEWORK

We assume that there are no special tuples such as punctuations [15] or watermarks [22] in input streams to explicitly indicate the progress of a stream with respect to the tuple timestamp, which is indeed the common case in real-world scenarios. Hence, we adopt the buffer-based disorder handling approach. However, in contrast to existing buffer-based approaches, we do not aim at maximizing the result quality, i.e., $\gamma(P)$. Instead, our objective is to provide a user-configurable result-latency vs. result-quality tradeoff, by allowing a user to specify his requirement $\Gamma$ on the minimum achieved recall. We then try to minimize the buffer-size, thus the result latency, under this user-specified recall requirement. In this section, we give an overview of our solution. Important notations are summarized in Table I.

### A. Overview

Fig. 2 describes the overall design of our quality-driven disorder handling framework for MSWJs. For the moment, let us assume an *MJoin*-style [5] join implementation, where the join operation is conducted by a single join operator which takes $m$ input streams directly, rather than by a tree of binary join operators. We discuss how to relax this assumption in Sec. V. In general, we follow a *two-step*, *prior-operation* disorder handling approach, i.e., handling first the intra-stream disorder and then the inter-stream disorder within input streams before presenting them to the join operator.

For each input stream, we use the $K$-slack algorithm [12], [23] to handle its intra-stream disorder. The basic idea of

$K$-slack is as follows: Taking stream $S_i$ ($i \in [1, m]$) as an example, a buffer of $K_i$ time units is used to sort tuples from $S_i$. Each time the local current time $^iT$ of $S_i$ (cf. Sec. II-A) is updated by a newly arrived tuple, every tuple $e_i$ in the buffer that satisfies the condition $e_i.ts + K_i \leq {}^iT$ is emitted. All such tuples are emitted in the timestamp order. An example of disorder handling using $K$-slack, where $K = 1$, is given in Fig. 3. According to the definition of the delay of a tuple (cf. Sec. II-A), the out-of-order tuple $e_{i,7}$ in this example has a delay of 2 time units. Because the buffer size $K$ is 1 time unit, $e_{i,7}$ cannot be sorted correctly, and appears still as an out-of-order tuple in the output of $K$-slack. However, the delay of $e_{i,7}$ in the output stream is reduced to 1 time unit. This example implies that, to successfully re-order a tuple with a delay of $k$ time units, we need a $K$-slack buffer of at least $k$ time units.

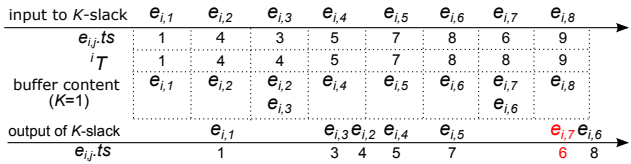

Fig. 3. Example of using $K$-slack to handle intra-stream disorder.

Existing work that uses $K$-slack for disorder handling either sets $K$ to a fixed value [14], [23], or increases $K$ dynamically to be equal to the maximum delay among so-far-observed out-of-order tuples [12]. In contrast, in this work, we dynamically adjust the value of $K$ in each $K$-slack component using a *Buffer-Size Manager*, to minimize the incurred result latency, while honoring the user-specified recall requirement $\Gamma$. The *Buffer-Size Manager* adapts the value of $K$ every $L$ time units. $L$ is a system parameter and is referred to as the *adaptation interval*. The value of $L$ is set such that it does not exceed the user-specified result-quality measurement period $P$, i.e., $L \leq P$. At the end of each adaptation interval, the value of $K$ in each $K$-slack component is determined for the next $L$ time units, based on the information provided by three other components: the *Statistics Manager*, the *Tuple-Productivity Profiler*, and the *Result-Size Monitor*. The *Statistics Manager* monitors input streams to estimate the tuple-delay distribution within each stream. These tuple-delay distributions play a critical role in our analytical model for estimating the recall produced under a certain configuration of $K$ in all $K$-slack components (cf. Sec. IV-A). The *Tuple-Productivity Profiler* interacts with the join operator to monitor the productivity of an input tuple. The objective is to learn the potential correlation between the delay and the productivity of input tuples, which is another important element in our analytical model (cf. Sec. IV-B). For brevity, we refer to this correlation as *DPcorr* hereafter. Finally, the *Result-Size Monitor* maintains a sliding window of $P-L$ time units on the stream of produced result tuples. Based on the produced result size, the *Buffer-Size Manager* calibrates the recall requirement used in a single adaptation step. The calibration is based on the following intuition: If the recall of the join results produced within the last $P - L$ time units is much higher than the user-specified requirement $\Gamma$, then a lower recall requirement can be used for the join results produced in the next adaptation interval, which implies that smaller $K$-slack buffers can be used in the next adaptation interval; and vice versa (Sec. IV-C).

Output streams of all $K$-slack components are then synchronized with each other by a *Synchronizer* using Alg. 1. Specifically, the *Synchronizer* maintains a buffer to sort arrived tuples, and a variable $T^{sync}$ to track the maximum timestamp among tuples that have left the buffer. Although merging multiple timestamp-ordered streams into a single timestamp-ordered stream has been discussed in existing work like [24], here, each input stream of the *Synchronizer* could still contain out-of-order tuples. As a result, a tuple $e$ arrived at the *Synchronizer* can be processed in two different ways. If $e.ts > T^{sync}$, then $e$ is inserted into the buffer, and every tuple that has the smallest timestamp is emitted from the buffer, as long as the buffer contains at least one tuple from each input stream (lines 4–8); otherwise, the tuple $e$ is emitted immediately (lines 9–10).

**Algorithm 1** Synchronize the output of all $K$-slack components
1: $T^{sync} \leftarrow 0$
2: $SyncBuf \leftarrow \emptyset$
3: **for** each tuple $e$ arrived at the *Synchronizer* **do**
4:     **if** $e.ts > T^{sync}$ **then**
5:         $SyncBuf$.insert($e$)
6:         **while** $SyncBuf$ has at least one tuple of each stream **do**
7:             $T^{sync} \leftarrow min\{e'.ts \mid e' \in SyncBuf\}$
8:             Emit every $e'$ having $e'.ts = T^{sync}$ from $SyncBuf$
9:     **else**
10:        Emit $e$ immediately

**Algorithm 2** MSWJ execution over the output of the *Synchronizer*
1: $^{\bowtie}T \leftarrow 0$
2: **for** each tuple $e_i$ ($i \in [1, m]$) arrived at the join operator **do**
3:     **if** $e_i.ts \geq {}^{\bowtie}T$ **then**
4:         $^{\bowtie}T \leftarrow e_i.ts$
5:         **for** window on each stream $S_j$ ($j \neq i$) **do**
6:             Remove tuples $e_j$ satisfying $e_j.ts < e_i.ts - W_j$
7:         Probe windows on each stream $S_j$ ($j \neq i$) and produce result tuples based on the join condition $p^{\bowtie}$
8:         Insert $e_i$ into the window on $S_i$
9:     **else if** $e_i.ts > {}^{\bowtie}T - W_i$ **then**
10:        Insert $e_i$ into the window on $S_i$
11:     Invoke the *Tuple-Productivity Profiler* to record the (estimated) productivity of $e_i$

The output stream of the *Synchronizer* is then processed by the MSWJ operator, whose basic idea is described in Alg. 2. Because of lines 9–10 in Alg. 1, the input to the join operator still contains out-of-order tuples. The join operator can detect these out-of-order tuples by using a variable $^{\bowtie}T$ to track the maximum timestamp among already received tuples. A received tuple $e_i$ ($i \in [1, m]$) is out of order if $e_i.ts < {}^{\bowtie}T$. For each received tuple $e_i$, if $e_i$ is an in-order tuple, then $^{\bowtie}T$ is updated if needed, and $e_i$ is processed following a three-step procedure: ① Invalidate expired tuples in windows on all other streams (lines 5–6). ② Join $e_i$ with remaining tuples in all other windows, and produce result tuples based on the given join condition (line 7). The timestamp assigned to each result tuple is $e_i.ts$. ③ Insert $e_i$ into the window on $S_i$ (line 8). If the received tuple $e_i$ is an out-of-order tuple, then steps ① and ② are skipped; hence, (a fraction of) result tuples that can be derived from $e_i$ are lost. However, if $e_i$ still falls into the current scope of the window on $S_i$ (i.e., $e_i.ts \geq {}^{\bowtie}T - W_i$), then $e_i$ could still contribute to deriving future result tuples. Hence, in this case, $e_i$ is still inserted into the window on $S_i$ (lines 9–10). Finally, the join operator invokes the *Tuple-Productivity Profiler* to record the productivity of $e_i$. The productivity of an out-of-order tuple is estimated based on past join results (cf. Sec. IV-B).

*B. The Same-K Policy*

Before describing the adaptation method applied by the *Buffer-Size Manager* in detail, we first introduce one general policy that the *Buffer-Size Manager* follows in each adaptation step. That is, all $K$-slack components in Fig. 2 use the same setting of $K$. We term this policy as the *Same-K* policy, which is asserted by Theorem 1.

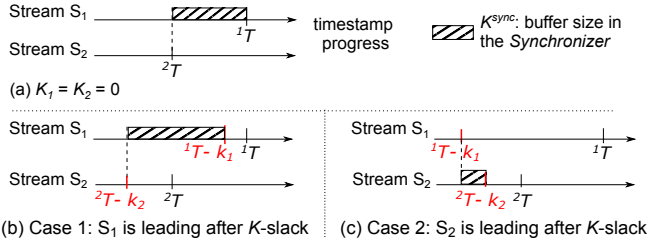

Fig. 4. Illustrative proof of the *Same-K* policy for 2-way joins.

**Theorem 1.** *With a two-step, buffer-based, prior-operation disorder handling approach (i.e., intra-stream disorder handling followed by inter-stream disorder handling before the join operation), for any buffer size configuration of the intra-stream disorder handling components (i.e., the $K$-slack components), where $K_1 = k_1$, $K_2 = k_2$, ..., $K_m = k_m$ ($k_i$ is a constant and $k_i \geq 0$), the overall effect of the disorder handling under this configuration, and thus the produced join output, is the same as that under the configuration $K_1 = K_2 = \cdots = K_m = k$, where $k = min\{^iT|i \in [1,m]\} - min\{^iT - k_i|i \in [1,m]\}$.*

Basically, Theorem 1 says that, independent of the intra-stream disorder characteristics within the input streams, we could always find a configuration $C$ where all $K$-slack components in the framework apply the same buffer size, to replace another configuration $C'$ where the $K$-slack components apply different buffer sizes, such that the join output produced under $C$ is the same as that produced under $C'$. Hence, it suffices to use the same value of $K$ in all $K$-slack components. In the following, we first illustratively show that Theorem 1 is true for 2-way joins, and then prove it for general $m$-way joins.

Assume a 2-way join with input streams $S_1$ and $S_2$, whose progress with respect to the tuple timestamp is as shown in Fig. 4a. $S_1$ is leading in terms of the timestamp progress whereas $S_2$ is lagging. It is important to note that, with Alg. 1, even if we do not handle the intra-stream disorder using $K$-slack components, i.e., $K_1 = K_2 = 0$, the *Synchronizer* would handle it, at least partially, for the leading stream. According to Alg. 1, we can deduce that, for $S_1$ and $S_2$ in Fig. 4a, the variable $T^{sync}$ in Alg. 1 equals to $^2T$; all so-far-arrived $S_1$ tuples, whose timestamps are within $(T^{sync}, {}^1T]$, are kept in and sorted by the synchronization buffer. Hence, the intra-stream disorder of $S_1$ is implicitly handled by a buffer of $^1T - T^{sync} = {}^1T - {}^2T$ time units. We denote this implicit buffer size within the *Synchronizer* that contributes to handling the intra-stream disorder of a stream $S_i$ by $K_i^{sync}$.

Assume that the $K$-slack components are configured with $K_1 = k_1$ and $K_2 = k_2$, and at least one of $k_1$ and $k_2$ is not zero. Then there are two possible cases[5]:

- *Case 1: $S_1$ remains leading after K-slack (Fig. 4b).* For stream $S_2$, the total buffer size for handling its intra-stream disorder is $k_2$. For stream $S_1$, in addition to the $K$-slack buffer, the synchronization buffer can further handle its intra-stream disorder, and $K_1^{sync} = (^1T - k_1) - (^2T - k_2)$. Hence, the total buffer size for handling the intra-stream disorder of $S_1$ is $k_1 + K_1^{sync} = {}^1T - {}^2T + k_2$. Compared to the case in Fig. 4a, for both streams, the total buffer size for intra-stream disorder handling is increased by $k_2$, which is equivalent to having a configuration $K_1 = K_2 = k_2$.

- *Case 2: $S_2$ becomes leading after K-slack (Fig. 4c).* In this case, the synchronization buffer keeps and sorts tuples from $S_2$ and $K_1^{sync} = 0$, $K_2^{sync} = (^2T - k_2) - (^1T - k_1)$. The total buffer size for handling intra-stream disorder is $k_1$ for $S_1$, and $k_2 + K_2^{sync} = {}^2T - {}^1T + k_1$ for $S_2$. Let $k = {}^2T - {}^1T + k_1$. Compared to the case in Fig. 4a, the total buffer size for handling the intra-stream disorder is increased by $k$ for both streams, which is equivalent to having a configuration $K_1 = K_2 = k$.

We now prove Theorem 1 for general MSWJs:

*Proof:* The nature of Alg. 1 determines that the variable $T^{sync}$ always equals to the maximum timestamp among tuples in the slowest stream produced by the $K$-slack components, where the meaning of "slow" was defined in Sec. II-A. If all $K$-slack components are configured with a buffer of size 0, then $T^{sync} = min\{^iT|i \in [1,m]\}$, and $K_i^{sync}$ for each stream $S_i$ can be determined as $K_i^{sync} = {}^iT - T^{sync} = {}^iT - min\{^iT|i \in [1,m]\}$. $K_i^{sync}$ is also the total buffer size for handling the intra-stream disorder of $S_i$ under this configuration. If the configuration for the $K$-slack components is $K_1 = k_1$, $K_2 = k_2$, ..., $K_m = k_m$ ($k_i \geq 0$), then $T^{sync} = min\{^iT - k_i|i \in [1,m]\}$, and $K_i^{sync} = (^iT - k_i) - T^{sync} = {}^iT - k_i - min\{^iT - k_i|i \in [1,m]\}$. Now the total buffer size for handling the intra-stream disorder of $S_i$ is $k_i + K_i^{sync} = {}^iT - min\{^iT - k_i|i \in [1,m]\}$. Compared to the case where $K_i = 0$ for all $i \in [1,m]$, for each stream, the total buffer size for handling its intra-stream disorder is increased by $(^iT - min\{^iT - k_i|i \in [1,m]\}) - (^iT - min\{^iT|i \in [1,m]\}) = min\{^iT|i \in [1,m]\} - min\{^iT - k_i|i \in [1,m]\}$. Hence, it is equivalent to having a configuration $K_1 = K_2 = \ldots = K_m = min\{^iT|i \in [1,m]\} - min\{^iT - k_i|i \in [1,m]\}$. ∎

The *Same-K* policy has another important benefit:

**Proposition 1.** *With the Same-K policy, for any two input streams $S_i$ and $S_j$, the time skew (cf. Sec. II-A) between their corresponding K-slack output streams is the same as the time skew between $S_i$ and $S_j$.*

*Proof:* $\forall i, j \in [1,m], \forall k, |(^iT - k) - (^jT - k)| = |^iT - {}^jT|$. ∎

From the discussion above, we can see that $K_i^{sync}$ for any stream is essentially the time skew between its corresponding $K$-slack output stream and the slowest output stream among all $K$-slack components. Proposition 1 suggests that, with the *Same-K* policy, we can determine $K_i^{sync}$ directly based on time skews among all raw input streams, regardless of the specific value of $K$ chosen for the $K$-slack components. We use the *Statistics Manager* in Fig. 2 to monitor $K_i^{sync}$ at runtime.

## IV. MODEL-BASED BUFFER-SIZE ADAPTATION

Thanks to the *Same-K* policy introduced in Sec. III-B, at each adaptation step, the *Buffer-Size Manager* only needs to determine a common $K$ value for all $K$-slack components in the framework. Recall that the objective is to find the minimum possible $K$ value to meet the user-specified result-quality (i.e., recall) requirement. To this end, in this section, we propose a model-based approach.

Specifically, at each adaptation step, we analytically model the recall of the join results that would be produced in the next adaptation interval, whose length is $L$, as a function

---
[5]The case in which both streams have the same timestamp progress after $K$-slack can be viewed as a special instance of either *Case 1* or *Case 2*.

**Algorithm 3** Model-based $K$ Adaptation
**Input:**
  System parameters:
    $L$ - Adaptation interval applied in the *Buffer-Size Manager*
    $b$ - Size of a basic window
    $g$ - $K$-search granularity
  Runtime statistics:
    Tuple-delay statistics maintained by *Statistics Manager*
    Tuple-productivity statistics maintained by *Tuple-Productivity Profiler*
    Result-size statistics maintained by *Result-Size Monitor*
**Output:** $k^*$ ($K$-slack buffer size applied in the next adaptation interval)
1: $MaxD^H \leftarrow$ current maximum tuple delay
2: $\Gamma' \leftarrow$ instant recall requirement derived from $\Gamma$
3: $k^* \leftarrow 0$
4: **while** ($k^* \leq MaxD^H \wedge \gamma(L, k^*) < \Gamma'$) **do**
5:   $k^* \leftarrow k^* + g$
  **return** $k^*$

---

of $K$, denoted by $\gamma(K, L)$. The model of $\gamma(K, L)$ is built using statistics about the tuple-delay distribution in each input stream, which are maintained by the *Statistics Manager* by monitoring a recent history of each stream, and tuple-productivity statistics, which are maintained by the *Tuple-Productivity Profiler*. We derive a requirement $\Gamma'$ for the recall measured over $L$ from the user-specified requirement $\Gamma$ for the recall measured over the result-quality measurement period $P$ (cf. Sec. II-B), and refer to $\Gamma'$ as the *instant recall requirement*. We then search for the minimum possible value of $K$ such that $\gamma(L, K) \geq \Gamma'$ holds. We denote this $K$ value by $k^*$.

Alg. 3 depicts the main algorithm for finding $k^*$ for the next adaptation interval. Currently, we search for $k^*$ using a *trial and error* method. Specifically, let $MaxD^H$ denote the current maximum tuple delay within a recent history of input streams that is monitored by the *Statistics Manager*. We examine $k^* = 0, k^* = g, k^* = 2g, k^* = 3g, \ldots$ ($g > 0$), until either $k^* > MaxD^H$ or $\gamma(L, k^*) \geq \Gamma'$; the parameter $g$ is referred to as the *K-search granularity*. Studying other algorithms for searching for $k^*$ is a topic of future work.

In the following, we give a detailed description of how $\gamma(L, K)$ is computed on each iteration in Alg. 3 in Sec. IV-A and Sec. IV-B, and describe how $\Gamma'$ is derived in Sec. IV-C.

### A. Modeling $\gamma(L, K)$

To estimate the recall within $L$ time units under a certain value of $K$, we need to estimate the result size that would be produced within $L$ under $K$, denoted by $N^{\bowtie}_{prod}(L, K)$, and the true result size within $L$, denoted by $N^{\bowtie}_{true}(L)$, if input streams are in order and synchronized with each other. $N^{\bowtie}_{true}(L)$ is independent of the setting of $K$.

**Estimating** $N^{\bowtie}_{true}(L)$. For an MSWJ with an arbitrary join condition $p^{\bowtie}$, $N^{\bowtie}_{true}(L)$ can be estimated by multiplying the result size of the corresponding *cross-join*, denoted by $N^{\times}_{true}(L)$, with the selectivity of the join condition, denoted by $sel^{\bowtie}$. We defer the discussion of estimating $sel^{\bowtie}$ to Sec. IV-B. For the moment, let us assume that $sel^{\bowtie}$ is known. $N^{\times}_{true}(L)$ is the sum of the number of cross-join results that would be produced at the arrival of each tuple $e_i$ ($i \in [1, m]$) during $L$ when input streams present no disorder. The number of cross-join results produced at the arrival of $e_i$ is a simple product of the number of tuples within the current window on every other stream $S_j$ ($j \neq i$). Let $S_i[W_i]$ denote the set of tuples that are within the current window on stream $S_i$, and $|S_i[W_i]|$ denote its cardinality. $|S_i[W_i]|$ can be estimated based on the average data arrival rate of $S_i$, denoted by $r_i$, and the window size $W_i$, i.e., $|S_i[W_i]| = r_i \cdot W_i$. For each stream $S_i$, the total number of tuples that would arrive during $L$ can be estimated based on $r_i$ as well, as $r_i \cdot L$. In summary, we estimate $N^{\bowtie}_{true}(L)$ as

$$N^{\bowtie}_{true}(L) = sel^{\bowtie} \cdot N^{\times}_{true}(L) = sel^{\bowtie} \cdot \sum_{i=1}^{m}(r_i \cdot L \cdot \prod_{j=1, j\neq i}^{m} r_j \cdot W_j)$$

$$= sel^{\bowtie} \cdot (\prod_{i=1}^{m} r_i) \cdot L \cdot (\sum_{i=1}^{m} \prod_{j=1, j\neq i}^{m} W_j). \quad (1)$$

**Estimating** $N^{\bowtie}_{prod}(L, K)$. We estimate $N^{\bowtie}_{prod}(L, K)$ again based on the result size of the corresponding cross-join under $K$, denoted by $N^{\times}_{prod}(L, K)$, and the join selectivity under $K$, denoted by $sel^{\bowtie}(K)$. We estimate $N^{\times}_{prod}(L, K)$ based on the following observations: (1) The join operator produces result tuples, if any, only at the arrival of in-order tuples (cf. Alg. 2). (2) When an in-order tuple $e_i$ arrives at the join operator, $S_j[W_j]$ ($j \neq i$) may be *incomplete*; because some tuples $e_j$ that satisfy $e_j.ts \geq e_i.ts - W_j$ may have not arrived because of the incomplete sorting by the $K$-slack components and the *Synchronizer*. We refer to such a tuple as a *missing tuple of* $S_j[W_j]$. (3) If we divide a window into small segments, then a recent segment of a window (i.e., a segment whose scope is closer to $e_i.ts$) often has more missing tuples than an older segment of the window. The reason is that, by the time $e_i$ arrives, out-of-order tuples whose timestamps fall into the scope of an older segment of a window might have already arrived at the join operator, and been inserted into the window (lines 9–10 in Alg. 2); whereas out-of-order tuples whose timestamps fall into the scope of a recent segment of the window can be observed only at later points in time.

These observations suggest that, to estimate $N^{\times}_{prod}(L, K)$, we need to estimate, under the given setting of $K$, the number of in-order tuples that the join operator would receive during $L$, and the degree of completeness of $S_j[W_j]$ ($j \neq i$) at the arrival of an in-order tuple $e_i$ ($i \in [1, m]$). To this end, we introduce a discrete random variable $D_i$ for each input stream $S_i$ to represent the coarse-grained delay of a tuple $e_i$ in $S_i$. Specifically, $D_i$ takes the value 0 if $delay(e_i) = 0$, the value 1 if $delay(e_i) \in (0, g]$, the value 2 if $delay(e_i) \in (g, 2g]$, and so forth; namely, the granularity for non-zero tuple delays is the same as the $K$-search granularity in Alg 3. Let $f_{D_i}$ denote the probability density function (pdf) of $D_i$, i.e., $f_{D_i}(d) = Pr[D_i = d]$, where $d \in \{0, 1, 2, \ldots\}$. Based on the assumption that the near future resembles the recent past, for each input stream $S_i$, the *Statistics Manager* monitors delays of input tuples that are within a window $R_i^{stat}$ over $S_i$'s recent history, and approximates $f_{D_i}$ using a histogram $H_i$. A fixed size for $R_i^{stat}$ is hard to define without a priori knowledge of the disorder pattern within the stream, and is also sensitive to changes in the disorder pattern. Hence, in this work, we use the adaptive window approach proposed in [25] to dynamically adjust the size of $R_i^{stat}$, based on the rate of changes detected from the delays of tuples in $R_i^{stat}$ itself. The size of each $R_i^{stat}$ ($i \in [1, m]$) is adjusted separately.

$f_{D_i}$ captures the tuple-delay characteristics within a raw input stream $S_i$. After the intra-stream disorder handling by the $K$-slack component, and the inter-stream disorder handling by the *Synchronizer*, the tuple-delay characteristics in the stream received by the join operator is different from $f_{D_i}$. Let $D_i^K$ denote a discrete random variable representing the coarse-grained delay of a tuple in the corresponding stream of $S_i$ that is received by the join operator under a certain $K$ setting. Let $f_{D_i^K}$ represent the pdf of $D_i^K$. We can capture the change from $f_{D_i}$ to $f_{D_i^K}$ based on the observation that, for any tuple $e_i$ in a raw input stream $S_i$, the delay of $e_i$ within the corresponding stream received by the join operator changes from $delay(e_i)$ to $delay^K(e_i)$, where $delay^K(e_i) = max\{0, delay(e_i) - K - K_i^{sync}\}$. Hence, we can derive $f_{D_i^K}$ from $f_{D_i}$ using Eq. (2).

$$f_{D_i^K}(d) = \begin{cases} \sum_{d'=0}^{(K+K_i^{sync})/g} f_{D_i}(d'), & d = 0 \\ f_{D_i}(d + \frac{K+K_i^{sync}}{g}), & d \in \{1, 2, 3, \dots\} \end{cases} \quad (2)$$

We assume that $K_i^{sync}$ for each stream keeps stable during the next adaptation interval, and estimate it as $\overline{K_i^{sync}} - min\{\overline{K_j^{sync}}|j \in [1, m]\}$, where $\overline{K_i^{sync}}$ represents the average of all $K_i^{sync}$ measurements obtained by the *Statistics Manager* within $R_i^{stat}$. Having estimated $K_i^{sync}$, we can then approximate $f_{D_i^K}$ using histogram $H_i$ as well. Moreover, using $f_{D_i^K}$, we can estimate the expected number of in-order $S_i$ tuples that would arrive at the join operator during $L$ as $r_i \cdot L \cdot f_{D_i^K}(0)$.

To capture the difference in the degree of completeness among different segments of a window $W_i$, we borrow the notion of *basic window* introduced in [3], and divide each window on a stream into basic windows of $b$ time units. The window on stream $S_i$ consists of $n_i = \lceil W_i/b \rceil$ basic windows. Let $w_i^l$ denote the $l$-th, $l \in [1, n_i]$, basic window on $S_i$; $w_i^1$ represents the most recent basic window. At any point in time, the scope of a basic window $w_i^l$ on $S_i$ can be determined as $(^{\bowtie}T - l \cdot b, ^{\bowtie}T - (l-1) \cdot b]$ for $l \in [1, n_i - 1]$, and $[^{\bowtie}T - W_i, ^{\bowtie}T - (n_i - 1) \cdot b]$ for $l = n_i$.

For $w_i^1$, among all $S_i$ tuples $e_i$ whose timestamps fall into the scope of $w_i^1$, tuples having $delay^K(e_i) = 0$ should have arrived. Hence, we estimate the expected number of tuples that are included in $w_i^1$, denoted by $|w_i^1|$, as $|w_i^1| = r_i \cdot b \cdot f_{D_i^K}(0)$. For $w_i^2$, all tuples having $delay^K(e_i) \in [0, \frac{b}{g}]$ should have arrived, and $|w_i^2|$ can be estimated as $r_i \cdot b \cdot \sum_{d=0}^{\frac{b}{g}} f_{D_i^K}(d)$. In general, $|w_i^l|$ for any $l \in [1, n_i]$ is estimated using Eq. (3).

$$|w_i^l| = \begin{cases} r_i \cdot b \cdot \sum_{d=0}^{\frac{(l-1)b}{g}} f_{D_i^K}(d), & l \in [1, n_i - 1] \\ r_i(W_i - (n_i - 1) \cdot b) \sum_{d=0}^{\frac{(n_i-1)b}{g}} f_{D_i^K}(d), & l = n_i \end{cases} \quad (3)$$

Based on estimations of $|w_i^l|$, we can estimate $|S_i[W_i]|$ as $\sum_{l=1}^{n_i} |w_i^l|$. Furthermore, $N_{prod}^{\bowtie}(L, K)$ can be estimated as

$$N_{prod}^{\bowtie}(L, K) = sel^{\bowtie}(K) \cdot \sum_{i=1}^{m} (r_i \cdot L \cdot f_{D_i^K}(0) \cdot (\prod_{j=1, j \neq i}^{m} \sum_{l=1}^{n_j} |w_j^l|))$$

$$= sel^{\bowtie}(K) \cdot (\prod_{i=1}^{m} r_i) \cdot L \cdot \sum_{i=1}^{m} (f_{D_i^K}(0) \prod_{j=1, j \neq i}^{m} \sum_{l=1}^{n_j} |w_j^l|). \quad (4)$$

Note that a bigger value of $b$ implies a more conservative estimation of $|S_i[W_i]|$ and thus $N_{prod}^{\bowtie}(L, K)$. When $b$ is chosen such that $n_i = 1$ for all $i \in [1, m]$, it means that the estimation of $|S_i[W_i]|$ considers only in-order tuples.

**Calculating** $\gamma(L, K)$. Having estimated $N_{true}^{\bowtie}(L)$ and $N_{prod}^{\bowtie}(L, K)$, we can calculate $\gamma(L, K)$ as

$$\gamma(L, K) = \frac{sel^{\bowtie}(K)}{sel^{\bowtie}} \cdot \frac{\sum_{i=1}^{m} (f_{D_i^K}(0) \prod_{j=1, j \neq i}^{m} \sum_{l=1}^{n_j} |w_j^l|)}{\sum_{i=1}^{m} \prod_{j=1, j \neq i}^{m} W_j}, \quad (5)$$

where the common factor $(\prod_{i=1}^{m} r_i) \cdot L$ in Eq. (1) and Eq. (4) is canceled off.

*B. Learning DPCorr and Estimating Join Selectivity*

We now discuss how to estimate $\frac{sel^{\bowtie}(K)}{sel^{\bowtie}}$ needed in Eq. (5). A naive strategy is to assume $sel^{\bowtie}(K) = sel^{\bowtie}$, which is equivalent to estimating $\gamma(L, K)$ based on result sizes of the corresponding cross-join. We denote this strategy by *EqSel*. However, when stream disorder is present and the disorder handling is incomplete, the streams received by the join operator are different from the streams in the ideal case, where all input streams are in order and synchronized. As a result, the join selectivity when stream disorder is present is often also different from the join selectivity in the ideal case. For instance, for the 2-way join in Fig. 5, where tuples are represented in the same way as in Fig. 1, when the input is ideal or the disorder handling is complete, the join selectivity is $\frac{1}{3}$ (Fig. 5a). If after disorder handling, a tuple in $S_1$ arrives at the join operator out of order, then the join selectivity is no longer $\frac{1}{3}$ (Fig. 5b and 5c). Hence, it is more reasonable to assume that $sel^{\bowtie}(K)$ is different from $sel^{\bowtie}$. We denote this strategy by *NonEqSel*.

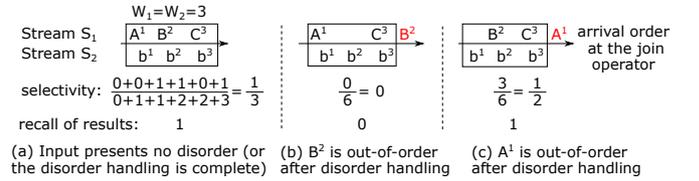

Fig. 5. Effect of out-of-order tuples arriving at the join operator on the join selectivity and the recall of results.

To estimate $sel^{\bowtie}(K)$ for different settings of $K$, we need to consider the correlation between the delay and the productivity of tuples, i.e., *DPcorr*; because, as implied by Fig. 5b and 5c, tuples with different delays do not necessarily have the same productivity, and the unsuccessful handling of an out-of-order tuple having a high productivity has a higher impact on the produced recall than the unsuccessful handling of an out-of-order tuple having a low productivity.

Extensive work (e.g., [26], [27]) exists, which uses synopsis of input streams (e.g., histograms, sketches, samples) to estimate the join result size or tuple productivities, and thereby the join selectivity. However, such *input-based* approaches do not work for joins with complex conditions, e.g., conditions involving user-defined functions [28]. To support arbitrary join conditions and to be able to estimate $sel^{\bowtie}(K)$ for different settings of $K$ in each adaptation step, we adopt an *output-based* approach; namely, we learn *DPcorr* by monitoring the output of the join. Such output-based approaches were also applied in prior work like [3], [20], [29] for different purposes.

Specifically, for each input tuple $e_i$ ($i \in [1, m]$), the $K$-slack component for stream $S_i$ annotates $e_i$ with $delay(e_i)$. Tuple $e_i$ carries this delay annotation through the *Synchronizer*. If $e_i$ arrives in order at the join operator, then during the join processing, the join operator records both the number of cross-join result size, $n^{\times}(e_i)$, that $e_i$ would derive, and the number of join results, $n^{\bowtie}(e_i)$, that $e_i$ actually derived, given the current window content $S_j[W_j]$ ($j \neq i$) on the other streams. The three numbers $delay(e_i)$, $n^{\times}(e_i)$, and $n^{\bowtie}(e_i)$ are then provided to the *Tuple-Productivity Profiler*. The *Tuple-Productivity Profiler* uses two maps, $\mathcal{M}^{\times}$ and $\mathcal{M}^{\bowtie}$, to maintain accumulated $n^{\times}$ and $n^{\bowtie}$, respectively, for each coarse-grained tuple delay observed within the last adaptation interval. The applied granularity for non-zero tuple delays is again $g$, which is consistent with $K$-search granularity in Alg. 3. If $e_i$ arrives out of order at the join operator, no join processing is conducted for $e_i$. We then estimate $n^{\bowtie}(e_i)$ and $n^{\times}(e_i)$ as the maximum $n^{\bowtie}(e)$ and $n^{\times}(e)$, respectively, over all in-order tuples $e$ received in the last adaptation interval.

Let $\mathcal{M}^{\times}[d]$ ($\mathcal{M}^{\bowtie}[d]$) represent the value to which the coarse-grained delay value $d$ is mapped in $\mathcal{M}^{\times}$ ($\mathcal{M}^{\bowtie}$), $\mathcal{M}^{\times}[d] = \sum_{delay(e)=d} n^{\times}(e)$ ($\mathcal{M}^{\bowtie}[d] = \sum_{delay(e)=d} n^{\bowtie}(e)$). Assuming that the join selectivity in the next adaptation interval is the same as the join selectivity in the last adaptation interval, for each $K$ value examined in Alg. 3, we estimate $sel^{\bowtie}(K)$ as $(\sum_{d=0}^{K} \mathcal{M}^{\bowtie}[d])/(\sum_{d=0}^{K} \mathcal{M}^{\times}[d])$.

The estimation of $sel^{\bowtie}$—the true join selectivity within $L$ when input streams present no disorder—is based on the following observation: For the join operator, the case where raw input streams present no disorder is equivalent to the case where input streams present disorder but the disorder is handled completely by using large-enough $K$-slack buffers. The join selectivities in both cases are the same. Therefore, we estimate $sel^{\bowtie}$ as $(\sum_{d=0}^{MaxD^{\mathcal{M}}} \mathcal{M}^{\bowtie}[d])/(\sum_{d=0}^{MaxD^{\mathcal{M}}} \mathcal{M}^{\times}[d])$, where $MaxD^{\mathcal{M}}$ represents the current maximum tuple delay in $\mathcal{M}^{\bowtie}$ (or $\mathcal{M}^{\times}$); because with a buffer of size $MaxD^{\mathcal{M}}$, any tuple $e$ having $delay(e) \leq MaxD^{\mathcal{M}}$ can be re-ordered correctly (cf. Sec. III-A). In summary, the estimation for $\frac{sel^{\bowtie}(K)}{sel^{\bowtie}}$ is

$$\frac{sel^{\bowtie}(K)}{sel^{\bowtie}} = \frac{\sum_{d=0}^{K} \mathcal{M}^{\bowtie}[d]}{\sum_{d=0}^{K} \mathcal{M}^{\times}[d]} \cdot \frac{\sum_{d=0}^{MaxD^{\mathcal{M}}} \mathcal{M}^{\times}[d]}{\sum_{d=0}^{MaxD^{\mathcal{M}}} \mathcal{M}^{\bowtie}[d]}. \quad (6)$$

### C. Deriving the Instant Recall Requirement

We now describe how the *instant recall requirement* $\Gamma'$ needed in Alg.3 is derived. This is done with the help of the runtime statistics maintained by the *Tuple-Productivity Profiler* and the *Result-Size Monitor* in Fig. 2.

Given the user-specified result-quality measurement period $P$, the *Result-Size Monitor* maintains the number of join results produced within the last $P - L$ time units, denoted by $N_{prod}^{\bowtie}(P-L)$. Let $N_{true}^{\bowtie}(P-L)$ denote the number of true join results within the last $P - L$ time units if input streams present no disorder. In general, to make the recall measured over $P$ at the end of the next adaptation interval meet the user-specified requirement $\Gamma$, the recall measured over the next adaptation interval, i.e., $\Gamma'$, should satisfy Eq. (7).

$$\frac{N_{prod}^{\bowtie}(P-L) + N_{true}^{\bowtie}(L) \cdot \Gamma'}{N_{true}^{\bowtie}(P-L) + N_{true}^{\bowtie}(L)} \geq \Gamma \quad (7)$$

Recall that we can estimate $N_{true}^{\bowtie}(L)$ by $\sum_{d=0}^{MaxD^{\mathcal{M}}} \mathcal{M}^{\bowtie}[d]$, where $\mathcal{M}^{\bowtie}$ maintains accumulated tuple productivities within the last adaptation interval. Furthermore, $N_{true}^{\bowtie}(P-L)$ can be estimated by summing up the $N_{true}^{\bowtie}(L)$ estimations obtained in the last $(P-L)/L$ adaptation intervals. Together with $N_{prod}^{\bowtie}(P-L)$ maintained by the *Result-Size Monitor*, we can derive $\Gamma'$ from Eq. (7). The final instant recall requirement applied in Alg. 3 is $max\{\Gamma', 1\}$.

### V. APPLICABILITY IN DISTRIBUTED MSWJ PROCESSING

MSWJs are by nature CPU and memory intensive. To support high volume input data, big window sizes, and expensive join conditions, scalable and distributed MSWJ processing has gained a lot of research interest recently (e.g., [6], [30]). An MSWJ can be implemented as either a single MJoin-style operator or a tree of binary join operators. Both types of implementation support distributed processing by splitting a macro $m$-way or binary join operator into smaller operator instances, exploiting *pipelined and data parallelism*.

As long as each operator instance follows the same processing semantics as in Alg. 2, then regardless of the specific implementation type, we can adapt the quality-driven disorder handling approach described in this paper to apply it in a distributed setup in the following way: Same as in Fig. 2, we use $K$-slack components to handle the intra-stream disorder of all input streams, and adjust the value of $K$ adaptively using the *Buffer-Size Manager*. Each input stream to an operator instance in the distributed setup is either the output stream of a $K$-slack component, or the output stream of another operator instance[6]. To deal with the inter-stream disorder among inputs, each operator instance is associated with a *Synchronizer*. Indeed, such a prior-join synchronization strategy was already applied in existing distributed join systems such as [6], [30].

Key information that the *Buffer-Size Manager* requires to make adaptation decisions include $f_{D_i}$, $K_i^{sync}$, $\mathcal{M}^{\times}$, $\mathcal{M}^{\bowtie}$, and $N_{true}^{\bowtie}(P-L)$; all other information can then be derived. We can obtain $f_D$, $K_i^{sync}$ by monitoring the raw input streams, and $N_{true}^{\bowtie}(P-L)$ by monitoring the final join output. To obtain $\mathcal{M}^{\bowtie}$, each operator instance needs to be instrumented so that, when receiving a tuple $e$ from a $K$-slack component, it annotates each intermediate result tuple produced at the arrival of $e$ with $delay(e)$; and when receiving such an annotated intermediate result tuple, it further propagates the tuple-delay annotation to each produced (intermediate) result tuple. Then $\mathcal{M}^{\bowtie}$ can be built by monitoring the final join output. To build $\mathcal{M}^{\times}$ accurately, for each $K$-slack output tuple $e_i$ ($i \in [1, m]$) that triggers the join processing at an operator instance, we need to know $|S_j[W_j]|$ for all $j$ ($j \neq i$). However, each $S_j[W_j]$ is often split into slices, which are maintained by different operator instances. Hence, obtaining accurate $|S_j[W_j]|$ would require communicating with all involved operator instances, which can be expensive. An alternative is to approximate each $|S_j[W_j]|$ using the average data rate $r_i$ monitored by the *Statistics Manager* and the window size $W_j$.

### VI. EXPERIMENTS

We implemented the proposed model-based, adaptive, quality-driven disorder handling approach in a prototypical version of SAP ESP [31]. In this section, we evaluate the effectiveness of our approach for varying user-specified requirements, and study effects of important system parameters. All experiments were conducted on a HP Z620 workstation, which has 24 cores (2.9GHz per core) and 96 GB RAM.

**Datasets and Queries.** For the evaluation, we used one real-world and two synthetic datasets with two, three, and four input streams. A different join query was used for each dataset.

- The real-world soccer-game dataset $D_{real}^{\times 2}$ consists of two streams ($s_1$ and $s_2$) of player-position data, which was collected by sensors during a 23-minute soccer training game, one stream for each team [1]. We projected each original stream onto (ts, sID, xCoord, yCoord), where sID identifies players and the pair of coordinates (xCoord, yCoord) encodes positions in the field. Each stream contains approximately 450k tuples. The maximum tuple delay is 22 seconds in $s_1$ and 26 seconds in $s_1$. The join query $Q^{\times 2}$ evaluated on $D_{real}^{\times 2}$ is to find all occurrences, within a 5-second sliding window, when the distance between two players, one from each team, is shorter than 5 meters. A custom function dist() calculates the distance and is the join condition for this scenario.

---
[6]The output of an operator instance is guaranteed to contain no intra-stream disorder because of the applied processing semantics.

$Q^{\times 2}$: SELECT * FROM S$_1$ [5 SEC], S$_2$ [5 SEC] WHERE
    dist(S$_1$.xCoord, S$_1$.yCoord, S$_2$.xCoord, S$_2$.yCoord) < 5

- The first synthetic dataset $D_{syn}^{\times 3}$ consists of three streams, which have the same schema (ts, a1). All streams start from a common timestamp $ts^{ini}$, which has millisecond (ms) granularity, and cover an interval of 30 minutes. For each stream S$_i$ ($i \in [1,3]$), tuples were generated sequentially as follows. Let $^i T = ts^{ini}$ initially. For each new tuple $e$, we increased $^i T$ by 10 ms (i.e., $^i T \mathrel{+}= 10$), and chose a random delay $delay(e)$ from $[0.0, 20.0]$ seconds using a Zipf distribution with skew $z_i^d$. We then set $e.ts$ to $^i T$ if $delay(e) = 0$, or $^i T - delay(e)$ otherwise. By increasing $^i T$ by 10 ms first for each newly generated tuple, we simulated a data rate of 100 tuples/sec. The applied Zipf skews $z_i^d$ for all streams were $z_1^d = 2.0$ and $z_2^d = z_3^d = 3.0$. The value of a1 for $e$ was generated randomly from the integer interval $[1, 100]$ using a Zipf distribution as well. To simulate a time-varying join selectivity, for each stream, the Zipf skew $z_i^{a1}$ for generating a1 was initialized to 1.0, and was changed, within the range $[0.0, 5.0]$, during the data generation. The time interval between two consecutive changes of $z_i^{a1}$ was chosen randomly from $[1, 10]$ minutes. The three streams are synchronized. We evaluated a 3-way join query on $D_{syn}^{\times 3}$:

$Q^{\times 3}$: SELECT * FROM S$_1$ [5 SEC], S$_2$ [5 SEC], S$_3$ [5 SEC]
    WHERE S$_1$.a1=S$_2$.a1 AND S$_2$.a1=S$_3$.a1

- The second synthetic dataset $D_{syn}^{\times 4}$ consists of four streams, whose schemas are S$_1$:(ts, a1, a2, a3), S$_2$:(ts, a1), S$_3$:(ts, a2), and S$_4$:(ts, a3). Timestamps and attribute values were generated in the same way and from the same domains as in $D_{syn}^{\times 3}$. The Zipf skew used for generating each attribute was also initialized with 1.0. Zipf skews used for generating tuple delays were $z_1^d = z_2^d = z_3^d = 3.0$ and $z_4^d = 4.0$. A 4-way join query was evaluated on $D_{syn}^{\times 4}$:

$Q^{\times 4}$: SELECT * FROM S$_1$ [3 SEC], S$_2$ [3 SEC],
           S$_3$ [3 SEC], S$_4$ [3 SEC]
    WHERE S$_1$.a1=S$_2$.a1 AND S$_1$.a2=S$_3$.a2 AND S$_1$.a3=S$_4$.a3

For each dataset, we generated a sorted version where tuples of all streams are globally ordered according to their timestamps. By evaluating $Q^{\times x}$ ($x \in \{2,3,4\}$) on the corresponding sorted dataset, we can obtain the true join results and therewith calculating the recall of results produced for the unsorted dataset.

**Default Parameter Configuration.** Unless otherwise stated, we used the following parameter configuration in our experiments: quality measurement period $P = 1\ min$, basic window size $b = 10\ ms$, $K$-search granularity $g = 10\ ms$, and adaptation interval $L = 1\ sec$.

**Metrics.** We consider the following metrics to evaluate our quality-driven disorder handling approach:

- *The average buffer size in a $K$-slack component.* Recall that all $K$-slack components share the same buffer size at any point in time (cf. Sec. III-B). The smaller the average $K$-slack buffer size, the lower the average result latency.
- *The percentage of recall measurements $\gamma(P)$ that fulfill the user-specified recall requirement $\Gamma$—in short, the req. fulfillment pct.* We denote this metric by $\Phi(\Gamma)$:

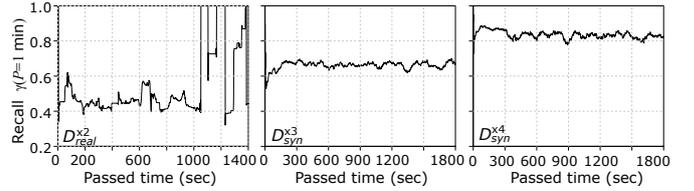

Fig. 6. Recall of join results produced by the *No-K-slack* baseline approach.

$$\Phi(\Gamma) = \frac{\#\ \gamma(P)\ \text{measurements that are not lower than}\ \Gamma}{\#\ \gamma(P)\ \text{measurements}}.$$

$\gamma(P)$ of join results produced under disorder handling was measured right before each adaptation of $K$. Recall measurements obtained during the first quality measurement period $P$ were excluded when computing $\Phi(\Gamma)$.

**Baseline Disorder Handling Approaches.** We consider two alternative *Buffer-Size Manager* (cf. Fig. 2) implementations as baselines, which manage buffer sizes in $K$-slack components in two extreme ways: (1) *No-K-slack*, which does not handle the intra-stream disorder of each stream via $K$-slack, i.e., $K_i = 0$ for all $i \in [1, m]$. (2) *Max-K-slack*, which updates the value of $K$ in each $K$-slack component dynamically to be equal to the maximum delay among so-far-observed tuples from all streams [12].

### A. Results of Baseline Disorder Handling Approaches

Fig. 6 shows recalls of join results $\gamma(P)$ produced by *No-K-slack* for each dataset, using default settings of $P$ and $L$. For $(D_{real}^{\times 2}, Q^{\times 2})$, the measured recall is only around 0.5 most of the time. The overall recall for $(D_{syn}^{\times 4}, Q^{\times 4})$ is the highest, but is only around 0.8, which is still a low result-quality for many applications. Fig. 6 implies that, to obtain a high result quality, doing inter-stream disorder handling only is not sufficient and intra-stream disorder handling is necessary.

TABLE II. RESULTS OF THE *Max-K-slack* BASELINE APPROACH.

|  | $(D_{real}^{\times 2}, Q^{\times 2})$ | $(D_{syn}^{\times 3}, Q^{\times 3})$ | $(D_{syn}^{\times 4}, Q^{\times 4})$ |
|---|---|---|---|
| Avg. $K$ (sec) | 19.96 | 19.72 | 13.88 |
| Avg. $\gamma(P)$ | 1.0 | 0.999 | 0.999 |

The average $K$ and $\gamma(P)$ produced by *Max-K-slack* are listed in Table II. The average $K$ for $(D_{syn}^{\times 4}, Q^{\times 4})$ is different from that for $(D_{syn}^{\times 3}, Q^{\times 3})$, because tuples with large delays appear later in streams of $D_{syn}^{\times 4}$ than in streams of $D_{syn}^{\times 3}$. The average $\gamma(P)$ is not always 1, because *Max-K-slack* does not guarantee complete disorder handling; each increase of $K$ in *Max-K-slack* is triggered by an out-of-order tuple $e$ whose delay is larger than the current $K$ value, and hence $e$ is still an out-of-order tuple in the output stream of $K$-slack.

Fig. 6 and Table II together corroborate that there is an inevitable tradeoff between the result latency and the result quality in disorder handling.

### B. Effectiveness Results

**Varying user-specified recall requirements.** We first studied the effectiveness of our approach under varying user-specified recall requirements $\Gamma$ for each pair of dataset and join query.

Recall that our approach is based on an analytical model of $\gamma(L, K)$, which uses statistics collected from past data to make predictions about future data (cf. Sec. IV). Due to the dynamic nature of data streams, it is impossible to make precise predictions; as a result, the derived buffer sizes may not guarantee a $\Phi(\Gamma)$ of 100%. However, a produced $\gamma(P)$ that violates $\Gamma$ can be indeed very close to $\Gamma$, and is acceptable in most scenarios. Hence, in addition to $\Phi(\Gamma)$, we also measured $\Phi(.99\Gamma)$, namely, the percentage of $\gamma(P)$ measurements that are not lower than $\Gamma$ by 1%. In this experiment, we also compared the two modeling strategies—*EqSel* and *NonEqSel*—that were discussed in Sec. IV-B.

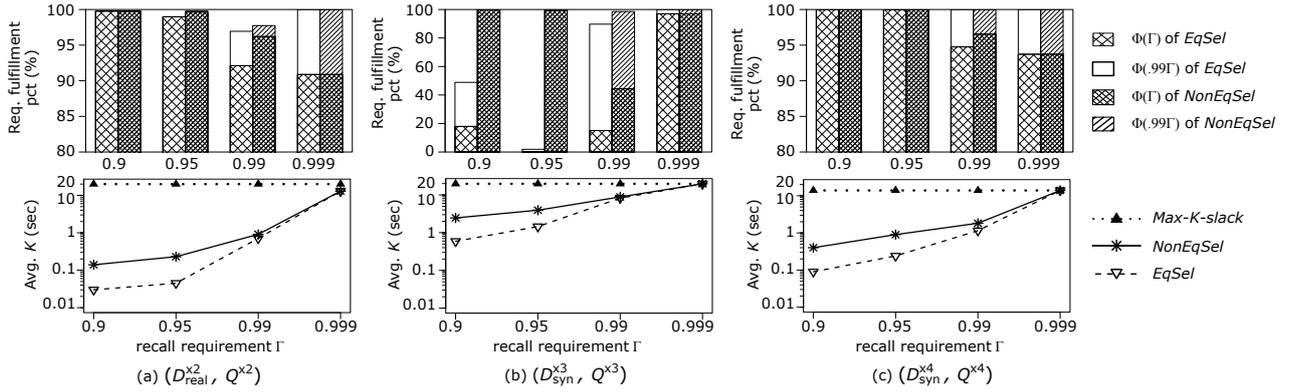

Fig. 7. Effectiveness of our quality-driven disorder handling approach for different data and join types under varying user-specified recall requirements $\Gamma$.

Fig. 7 shows the experimental results. For ease of comparison, we included the average $K$ produced by *Max-K-slack* for each dataset in the corresponding sub-figure. We can see that, for both *EqSel* and *NonEqSel*, the average $K$ goes up as $\Gamma$ increases, which again reveals the result-latency vs. result-quality tradeoff. *NonEqSel* produces a bit higher average $K$ than *EqSel*; the $\Phi(\Gamma)$ and $\Phi(.99\Gamma)$ produced by *NonEqSel* are not much higher than those produced by *EqSel* for $(D_{real}^{\times 2}, Q^{\times 2})$ and $(D_{syn}^{\times 4}, Q^{\times 4})$, but are significantly higher for $(D_{syn}^{\times 3}, Q^{\times 3})$. For each examined (dataset, query) pair, *NonEqSel* achieves a $\Phi(.99\Gamma)$ of at least 97% for all examined values of $\Gamma$. Hence, *NonEqSel* is more robust than *EqSel* towards different datasets and join queries, and is used in all following experiments.

Compared to the state-of-the-art approach *Max-K-slack*, our approach, using *NonEqSel*, can significantly reduce the average $K$, thus the result latency incurred by disorder handling, while still honoring the user-specified result-quality requirement. For instance, even for a high recall requirement $\Gamma = 0.99$, the average $K$ is reduced by more than 95% for $(D_{real}^{\times 2}, Q^{\times 2})$. For an even higher recall requirement $\Gamma = 0.999$, our approach still achieves a 35% reduction in the average $K$ for $(D_{real}^{\times 2}, Q^{\times 2})$, and reduces to the *Max-K-slack* approach for the other two datasets.

**Varying user-specified result-quality measurement periods.** We further studied the effectiveness of our approach under varying user-specified result-quality measurement periods $P$. Fig. 8 shows the results for $(D_{real}^{\times 2}, Q^{\times 2})$ and $(D_{syn}^{\times 3}, Q^{\times 3})$ under $\Gamma=0.95$ and $\Gamma=0.99$, although we observed similar results for $(D_{syn}^{\times 3}, Q^{\times 3})$ and other $\Gamma$ values. The other parameters took default values.

As expected, it is more difficult to obtain a high $\Phi(\Gamma)$ and $\Phi(.99\Gamma)$ for small values of $P$ than for big values of $P$; because the smaller the value of $P$, the slimmer the chance that a low recall of join results produced within one adaptation interval gets compensated by recalls produced in other adaptation intervals within the same measurement period. Nevertheless, we still achieve a $\Phi(.99\Gamma)$ of more than 90% for all examined values of $P$.

### C. Effect of the Adaptation Interval System Parameter

Fig. 9 studies the effect of the adaptation interval $L$ on the performance of our approach. where $L$ is varied from 0.1

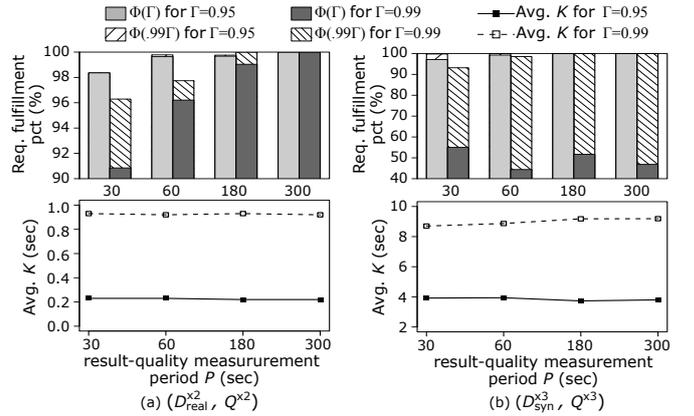

Fig. 8. Effectiveness of our quality-driven disorder handling approach under varying user-specified result-quality measurement periods $P$.

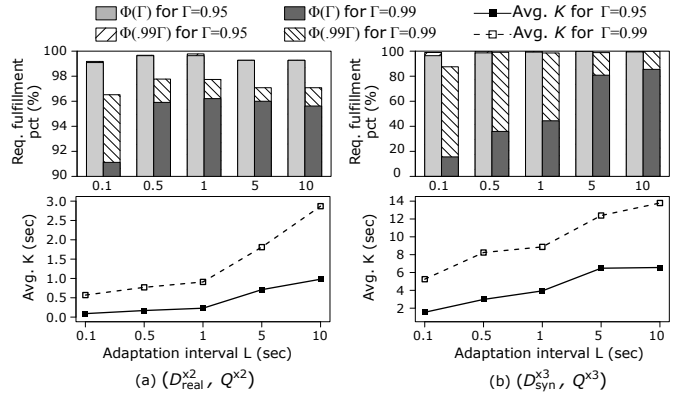

Fig. 9. Effect of the adaptation interval $L$ on the performance of our approach.

to 10 seconds. The figure reports results for $(D_{real}^{\times 2}, Q^{\times 2})$ and $(D_{syn}^{\times 3}, Q^{\times 3})$ under $\Gamma = 0.95$ and $\Gamma = 0.99$. The other parameters took default values.

The average $K$ grows noticeably as $L$ increases. This can be explained by the selectivity estimation done in Eq. (6). Recall that we estimate the productivity of an out-of-order tuple arrived at the join operator conservatively as the maximum tuple productivity observed within the last adaptation interval. The maximum tuple productivity observed within a long interval is often higher than that observed in a short interval; hence, the estimated selectivity is smaller, which often leads to a smaller result of Eq. (6). Moreover, for longer adaptation intervals, a large value of $K$ determined in one adaptation step is also applied for a longer time. The increase of $K$ leads to a great increase of $\Phi(\Gamma)$ for $(D_{syn}^{\times 3}, Q^{\times 3})$ under $\Gamma = 0.99$, but has little effect on the achieved result quality for the other three examined cases. We find that $L = 1$ second produces a good tradeoff between the average $K$ and the achieved result quality in our experiments.

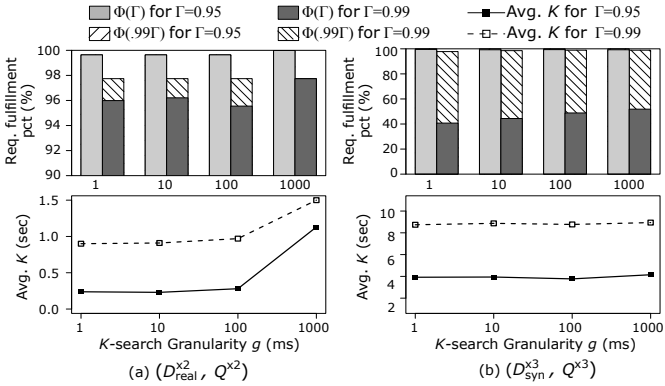

Fig. 10. Effect of the $K$-search granularity $g$ on the performance of our approach.

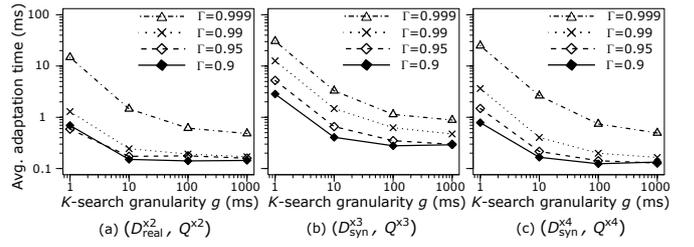

Fig. 11. Time needed to determine the optimal $K$ in an adaptation step.

### D. Effect of the $K$-Search Granularity System Parameter

Recall that in each adaptation step, we search for the optimal setting of $K$ by examining possible $K$ values incrementally, starting from the value 0 (cf. Alg. 3). The search granularity is $g$. Fig. 10 studies the effect of the setting of $g$ on the performance of our approach. The value of $g$ was varied from 1 to 1000 ms. The other parameters took default values. As $g$ increases, the average $K$ increases noticeably for $(D_{real}^{\times 2}, Q^{\times 2})$, but has nearly no changes for $(D_{syn}^{\times 3}, Q^{\times 3})$. This result implies that the value of $g$ has stronger effect in scenarios where satisfying the result-quality requirement $\Gamma$ requires a small buffer size than in scenarios where satisfying $\Gamma$ requires a big buffer size. We empirically chose $g = 10$ ms as the default setting in our system.

### E. Adaptation Overhead

Last, we studied the overhead of our model-based buffer size adaptation approach in terms of the time needed to find the optimal value of $K$ in an adaptation step; namely, the runtime of Alg. 3. The adaptation time is influenced by the number of input streams $m$, the user-specified recall requirement $\Gamma$, and the $K$-search granularity $g$. The number of input streams varies from 2 to 4 in our three datasets. For each dataset, we picked different combinations of $\Gamma$ and $g$ values, and ran the corresponding join query three times on that dataset for each $(\Gamma, g)$ combination. The other parameters took default values. For each $(\Gamma, g)$ combination, we recorded the average time of an adaptation step over all three runs of the query under that combination. The results are shown in Fig. 11. As expected, the adaptation time goes down as $g$ increases and $\Gamma$ decreases. For $g \geq 10$ ms, the average adaptation time is below 5 ms for the highest examined value of $\Gamma$ on all datasets. The average adaptation time for lower $\Gamma$ values is even smaller—below 1 ms. Moreover, in our implementation, the *Buffer-Size Manager* and the join operator run in separate threads; hence, the buffer-size adaptation time overlaps with the join processing time.

## VII. RELATED WORK

**MSWJ over Data Streams.** Representative early work on processing MSWJs over data streams includes [4], [5], [20]. Viglas et al. [5] proposed the MJoin algorithm and showed that, in many cases, MJoin can produce join results at a faster rate than any tree of binary join operators. The work of [4] compared the efficiency of nested-loop joins and hash-joins for different types of join conditions, and proposed join ordering heuristics based on a main-memory per-unit-time cost model. The work of [20] focused mainly on the problem of join ordering, and proposed an adaptive solution.

Aiming at a high throughput, several join algorithms (e.g., CellJoin [32], HELLS-join [33], and handshake join [11]) were proposed to exploit the high degree of parallelism in modern hardware such as multi-core CPU, GPU, and FPGA. The trend towards cloud computing has motivated much research on distributed join processing over a shared-nothing cluster. An early work along this line is [34]. It proposed a load diffusion operator, which can dynamically distribute the join workload via adaptive tuple routing, while preserving the MSWJ semantics. PSP [6] splits the states of an MSWJ into disjoint slices in the time domain and connects these state slices in a virtual ring architecture. BiStream [30] is a distributed join processing system based on a *join-biclique* model, which is proved to be more memory efficient than the *join-matrix* [35] model.

The work discussed above focused mainly on the efficiency of processing MSWJs, and all ignored the existence of intra-stream disorder. Some work like BiStream bypassed the intra-stream disorder problem by assigning timestamps to received input tuples based on the system time, which is not applicable for applications that require a join semantics that is based on the application time (i.e., based on tuples timestamps assigned at data sources). Most work also ignored the existence of inter-stream disorder, except the work of [10], [24], [30], which handled the inter-stream disorder by synchronizing input streams prior to the join operation. The work of [10] also compared the approach that enforces ordered processing of input and the approach that allows out-of-order processing of input but enforces ordered release of join results in terms of the memory consumption and the tuple response time for 2-way joins. Wu et al. [36] considered the presence of both intra-stream and inter-stream disorder. They also followed the two-step disorder handling approach. However, they did not focus on sliding window queries. Moreover, none of the aforementioned work considered quality-driven disorder handling.

**Disorder Handling Approaches.** Many disorder handling approaches have been proposed, especially for handling the intra-stream disorder. At a high level, they can be classified as follows: buffer-based approach (e.g., [12]), which uses buffers to reorder tuples, punctuation-based approach (e.g., [15]), which relies on special tuples—punctuations—embedded in input streams to communicate the stream progress, speculation-based approach (e.g., [37]), which produces results speculatively and applies a compensation technique to correct early emitted results when out-of-order arrivals are observed, and hybrid approach (e.g., [18]), which combines two of the former three approaches. We refer readers to our previous work [17] for a detailed discussion. Most existing work tries to either maximize the result quality or minimize the result latency. Work that attempts to balance these two requirements only considered controlling the tradeoff from the latency side, e.g., allowing a user to specify the desired buffer size for reordering input. In contrast, our work aims to control the tradeoff from the quality side.

**Load Shedding for Joins.** Our work is also closely related to load shedding in data stream processing, whose basic idea is to process a selected portion of input tuples when the available system resources (e.g., CPU, memory) cannot match the demand of the query. Existing work that focuses on join queries includes [3], [8], [19]. Result tuples produced under load shedding is only a subset of the true result tuples, and how to maximize the output rate under resource constraints is one of the key research topics. We can see that the effect of a dropped tuple during load shedding on the result quality is the same as that of an unsuccessfully handled out-of-order tuple during disorder handling. However, in load shedding, to determine which tuples to drop, we need to consider only the contribution of a tuple to the result quality based on its productivity; whereas in quality-driven disorder handling, to determine which out-of-order tuples could be handled unsuccessfully, we need to consider not only the productivity but also the delay of a tuple. Namely, one more dimension needs to be considered in quality-driven disorder handling.

## VIII. CONCLUSION

In this paper, we proposed a buffer-based, adaptive, quality-driven disorder handling framework for processing MSWJs over out-of-order and unsynchronized data streams. The framework is generic and supports MSWJs with arbitrary join conditions. Moreover, we proposed a novel model-based approach for adapting the buffer size at runtime to minimize the result latency incurred by disorder handling, while honoring user-specified result-quality requirements. The proposed model directly captures the relationship between the buffer size and the recall of produced join results, thereby allowing to search for the optimal buffer size at each adaptation step. Experimental results showed the effectiveness of our approach. Compared to the state of the art, we can significantly reduce the applied buffer size for disorder handling, thus the incurred result latency, while still providing the desired result quality.

## REFERENCES

[1] C. Mutschler, H. Ziekow, and Z. Jerzak, "The DEBS 2013 grand challenge," in *ACM DEBS*, 2013, pp. 289–294.
[2] B. Chandramouli, M. Ali, J. Goldstein, B. Sezgin, and B. Raman, "Data stream management systems for computational finance," *Computer*, vol. 43, no. 12, pp. 45–52, 2010.
[3] B. Gedik, K.-L. Wu, P. Yu, and L. Liu, "A load shedding framework and optimizations for m-way windowed stream joins," in *IEEE ICDE*, 2007, pp. 536–545.
[4] L. Golab and M. T. Özsu, "Processing sliding window multi-joins in continuous queries over data streams," in *VLDB*, 2003, pp. 500–511.
[5] S. D. Viglas, J. F. Naughton, and J. Burger, "Maximizing the output rate of multi-way join queries over streaming information sources," in *VLDB*, 2003, pp. 285–296.
[6] S. Wang and E. Rundensteiner, "Scalable stream join processing with expensive predicates: Workload distribution and adaptation by time-slicing," in *EDBT*, 2009, pp. 299–310.
[7] J. Kang, J. F. Naughton, and S. Viglas, "Evaluating window joins over unbounded streams." in *IEEE ICDE*, 2003, pp. 341–352.
[8] U. Srivastava and J. Widom, "Memory-limited execution of windowed stream joins," in *VLDB*, 2004, pp. 324–335.
[9] ——, "Flexible time management in data stream systems," in *ACM PODS*, 2004, pp. 263–274.
[10] M. A. Hammad, W. G. Aref, and A. K. Elmagarmid, "Optimizing in-order execution of continuous queries over streamed sensor data," in *SSDBM*, 2005, pp. 143–146.
[11] P. Roy, J. Teubner, and R. Gemulla, "Low-latency handshake join," *PVLDB*, vol. 7, no. 9, pp. 709–720, 2014.
[12] C. Mutschler and M. Philippsen, "Distributed low-latency out-of-order event processing for high data rate sensor streams," in *IEEE IPDPS*, 2013, pp. 1133–1144.
[13] B. Chandramouli, J. Goldstein, and D. Maier, "High-performance dynamic pattern matching over disordered streams," *PVLDB*, vol. 3, no. 1-2, pp. 220–231, 2010.
[14] D. J. Abadi, D. Carney, U. Çetintemel, M. Cherniack, C. Convey, S. Lee, M. Stonebraker, N. Tatbul, and S. Zdonik, "Aurora: A new model and architecture for data stream management," *VLDB Journal*, vol. 12, no. 2, pp. 120–139, 2003.
[15] J. Li, K. Tufte, V. Shkapenyuk, V. Papadimos, T. Johnson, and D. Maier, "Out-of-order processing: A new architecture for high-performance stream systems," *PVLDB*, vol. 1, no. 1, pp. 274–288, 2008.
[16] Y. Ji, H. Zhou, Z. Jerzak, A. Nica, G. Hackenbroich, and C. Fetzer, "Quality-driven continuous query execution over out-of-order data streams," in *ACM SIGMOD*, 2015, pp. 889–894.
[17] ——, "Quality-driven processing of sliding window aggregates over out-of-order data streams," in *ACM DEBS*, 2015, pp. 68–79.
[18] S. Krishnamurthy, M. J. Franklin, J. Davis, D. Farina, P. Golovko, A. Li, and N. Thombre, "Continuous analytics over discontinuous streams," in *ACM SIGMOD*, 2010, pp. 1081–1092.
[19] Y.-N. Law and C. Zaniolo, "Load shedding for window joins on multiple data streams," in *IEEE ICDE Workshop*, 2007, pp. 674–683.
[20] S. Babu, R. Motwani, K. Munagala, I. Nishizawa, and J. Widom, "Adaptive ordering of pipelined stream filters," in *ACM SIGMOD*, 2004, pp. 407–418.
[21] B. Babcock, S. Babu, M. Datar, R. Motwani, and J. Widom, "Models and issues in data stream systems," in *ACM PODS*, 2002, pp. 1–16.
[22] T. Akidau, A. Balikov, K. Bekiroğlu, S. Chernyak, J. Haberman, R. Lax, S. McVeety, D. Mills, P. Nordstrom, and S. Whittle, "MillWheel: Fault-tolerant stream processing at internet scale," *PVLDB*, vol. 6, no. 11, pp. 1033–1044, 2013.
[23] M. Li, M. Liu, L. Ding, E. Rundensteiner, and M. Mani, "Event stream processing with out-of-order data arrival," in *IEEE ICDCS Workshop*, 2007, pp. 67–67.
[24] V. Gulisano, R. Jimenez-Peris, M. Patino-Martinez, C. Soriente, and P. Valduriez, "StreamCloud: An elastic and scalable data streaming system," *IEEE TPDS*, vol. 23, no. 12, pp. 2351–2365, 2012.
[25] A. Bifet and R. Gavald, "Learning from time-changing data with adaptive windowing," in *SIAM SDM*, 2007, pp. 443–448.
[26] A. Swami and K. Schiefer, "On the estimation of join result sizes," in *EDBT*, 1994, vol. 779, pp. 287–300.
[27] F. Rusu and A. Dobra, "Sketches for size of join estimation," *ACM Trans. Database Syst.*, vol. 33, no. 3, pp. 15:1–15:46, 2008.
[28] S. Chaudhuri, "Query optimizers: time to rethink the contract?" in *ACM SIGMOD*, 2009, pp. 961–968.
[29] P.-A. Larson, W. Lehner, J. Zhou, and P. Zabback, "Cardinality estimation using sample views with quality assurance," in *ACM SIGMOD*, 2007, pp. 175–186.
[30] Q. Lin, B. C. Ooi, Z. Wang, and C. Yu, "Scalable distributed stream join processing," in *ACM SIGMOD*, 2015, pp. 811–825.
[31] SAP Event Stream Processor, "http://www.sap.com/pc/tech/database/software/sybase-complex-event-processing/index.html."
[32] B. Gedik, P. S. Yu, and R. R. Bordawekar, "Executing stream joins on the cell processor," in *VLDB*, 2007, pp. 363–374.
[33] T. Karnagel, D. Habich, B. Schlegel, and W. Lehner, "The HELLS-join: A heterogeneous stream join for extremely large windows," in *ACM DaMoN*, 2013, pp. 2:1–2:7.
[34] X. Gu, P. Yu, and H. Wang, "Adaptive load diffusion for multiway windowed stream joins," in *IEEE ICDE*, 2007, pp. 146–155.
[35] M. Elseidy, A. Elguindy, A. Vitorovic, and C. Koch, "Scalable and adaptive online joins," *PVLDB*, vol. 7, no. 6, pp. 441–452, 2014.
[36] J. Wu, K.-L. Tan, and Y. Zhou, "Window-oblivious join: A data-driven memory management scheme for stream join," in *SSBDM*, July 2007, pp. 21–21.
[37] M. Liu, M. Li, D. Golovnya, E. Rundensteiner, and K. Claypool, "Sequence pattern query processing over out-of-order event streams," in *IEEE ICDE*, 2009, pp. 784–795.